\documentclass{ws-procs975x65}
\usepackage{mathrsfs}

\newcommand{\re}[1]{(\ref{eq:#1})}

\def\phi{\varphi}

\def\rho{\varrho}
\def\d{\mathrm{d}}

\renewcommand{\vec}[1]{\boldsymbol{#1}}

\newcommand{\ssp}{\scriptscriptstyle{+}}
\newcommand{\ssm}{\scriptscriptstyle{-}}

\newcommand{\fp}{\mathcal{A}}

\newcommand{\rem}[1]{}

\begin{document}

\title{CLASSICAL LIMITS OF BOOST-ROTATION SYMMETRIC SPACETIMES}

\author{DAVID KOFRO\v{N}$^1$ and JI\v{R}\'I BI\v{c}\'AK$^2$}
\address{Institute of Theoretical Physics, Faculty of Mathematics and Physics,\\
Charles University, V Hole\v{s}ovi\v{c}k\'{a}ch 2, 180\,00 Prague 8, Czech Republic\\
$^1$E-mail: d.kofron@gmail.com\qquad $^2$E-mail: Jiri.Bicak@mff.cuni.cz\\
http://utf.mff.cuni.cz}

\bodymatter

\begin{abstract}
Boost-rotation symmetric spacetimes are exceptional as they are the only exact asymptotically flat solutions to the Einstein equations describing spatially bounded sources (``point-like'' particles, black holes) undergoing non-trivial motion (``uniform acceleration'') with radiation. We construct the Newtonian limit of these spacetimes: it yields fields of uniformly accelerated sources in classical mechanics. We also study the special-relativistic limit of the charged rotating C-metric and so find accelerating electromagnetic magic field.
\end{abstract}

\keywords{boost-rotation symmetry, C-metric, Newtonian limit}

\section{Introduction}

The boost-rotation (BR) symmetric solutions describe spatially bounded charged rotating objects undergoing a ``uniform acceleration''; they are asymptotically flat in the sense that they admit smooth though not complete null infinity. Moreover, they are radiative (it has been proven that boost Killing vector is the only one additional Killing vector to axial Killing vector which does not exclude gravitational radiation). They were used both in analytical and numerical relativity (for their brief review see Refs.\citelow{BiSchm,BiE}).

\section{Boost-rotation symmetric spacetimes and the C-metric} \label{chap:BRsC}

The electrovacuum BR symmetric metric\cite{BiSchm,BiPr}, generally of algebraic type I, in global coordinates reads
\begin{equation}
\d s^2 = \frac{ e^{\mu}\left[z\,\d t-t\,\d z - \left( z^2-t^2 \right)\omega\,\d\,\phi\right]^2-e^{\nu} \left(z\,\d z-t\,\d t\right)^2}{z^2-t^2}\ - e^{\nu}\,\d \rho^2-e^{-\mu}\rho^2\,\d \phi^2 \,,
\label{eq:BRmetric}
\end{equation}
where $\mu,\,\nu$ and $\omega$ are functions of $(\rho^2,\,t^2-z^2)$ only. Substituting \re{BRmetric} for non-rotating case ($\omega=0$) into vacuum Einstein's field equations we get the flat space wave equation $\square \mu = 0$. With non-vanishing rotation or electromagnetic field present system of coupled non-linear PDEs arises. Therefore, in general, only non-rotating vacuum solutions are explicitly known.

The charged rotating C-metric -- also a member of BR symmetric spacetimes, but algebraically special (of type D) --  reads\cite{HT2,BiKofAcc}
\begin{equation*}
\d s^2 = \frac{1}{A^2(x-y)^2}\,\biggl\{ \frac{\mathcal{G}(y)}{1+\left( aAxy \right)^2}\;\Bigl[ \left( 1+a^2A^2x^2 \right)K\d t+aA\left( 1-x^2 \right)K\d\phi \Bigr]^2
\end{equation*}
\begin{multline}
-\frac{1+\left( aAxy \right)^2}{\mathcal{G}(y)}\,\d y^2 +\frac{1+\left( aAxy \right)^2}{\mathcal{G}(x)}\,\d x^2\\
+\frac{\mathcal{G}(x)}{1+\left( aAxy \right)^2}\;\Bigl[ \left( 1+a^2A^2y^2 \right)K\d\phi+aA\left(y^2-1\right)K\d t \Bigr]^2\biggr\}\,,
\label{eq:rCMr}
\end{multline}
where the structure function $\mathcal{G}(\xi)$ following from the Einstein-Maxwell equations is
\begin{equation}
\mathcal{G}(\xi) = \left( 1-\xi^2 \right)\left( 1+r_{\ssp}A\xi \right)\left( 1+r_{\ssm}A\xi \right)\,\quad r_\pm=m \pm \sqrt{m^2-a^2-q^2}\,.
\label{eq:rG}
\end{equation}
Here $m$, $a$, $q$ and $A$ are respectively the mass, rotation, charge and acceleration parameters, $K$ is a constant scaling angular coordinate.
 
The explicit (but complicated) transformation of the C-metric \re{rCMr} with $q=a=0$ to the Weyl form and subsequently to the form of \re{BRmetric} can be found in Ref \citelow{HT1}.

The 4-potential of the electromagnetic field corresponding to \re{rCMr} is
\begin{equation}
\vec{\fp} = \frac{Kqy}{1+\left( aAxy \right)^2}\,\left[ \left(1+a^2A^2x^2\right)\vec{\d} t+aA\left( 1-x^2 \right)\vec{\d}\phi \right]\,.
\label{eq:rA}
\end{equation}

\section{The Newtonian limit of boost-rotation symmetric spacetimes}

In our work \cite{BiKofNL}, based upon the Ehlers frame theory \cite{Ehl-Els}, we found physically plausible Newtonian limits of the BR symmetric solutions. 

What has to be done is to introduce the causality constant $\lambda=c^{-2}$ in the metric and choose suitable coordinate system\footnote{or congruence of observers, as we used for charged C-metric}. Then in the limit $\lambda\rightarrow 0$ the Ehlers frame theory goes over to the Newton-Cartan theory where two distinct spatial and temporal metrics together with affine connection occur. The connection is then given by the Newtonian potential $\Phi$ as follows: $\Gamma^\alpha_{\beta\gamma} = t_{,\beta}\,t_{,\gamma}\,s^{\alpha\delta}\,\Phi_{,\delta}$.

The next key point is to observe that even in special relativity the worldline of a uniformly accelerated particle is hyperbola $z = \sqrt{c^4g^{-2}+c^2t^2} = \sqrt{\lambda^{-2}g^{-2}+t^2\lambda^{-1}}$. The whole hyperbola disappears to infinity in the limit $\lambda\rightarrow 0$ and the acceleration horizon (the ``roof'' -- see Ref.\citelow{BiSchm}) becomes the hyperplane $t=0$.
Hence, in order to obtain a nontrivial limit, we have to ``go'' to infinity with the particle. We do this by introducing a new coordinate $\zeta$ by 
\begin{equation}
z = \zeta + \frac{1}{\lambda g}\,,
\label{eq:subs}
\end{equation}
and so we make a $\lambda$-dependent shift of $z$. Then in the limit the hyperbola turns into the parabola. 
 
Before performing the limit $\lambda\rightarrow 0$ it is crucial to make the substitution \re{subs}. 
We can calculate the nontrivial components of affine connection and find 
\begin{equation}
\Gamma^a_{tt} = \lim_{\lambda\rightarrow 0}  \left[ \left. \left( \frac{1}{2} \frac{\mu_{,a}}{\lambda}\right)\right|_{z=\zeta+\frac{1}{\lambda g}} \right]\,, \qquad a=\{\rho,\,z\}\,.
\label{eq:CF}
\end{equation}
Notice that the other metric function $\nu$ does not enter the Newtonian results.

After long but straightforward calculations of the limit \re{CF}, we obtain the resulting Newtonian potential. Remarkably, complicated form (for functions $\mu,\,\nu$ entering \re{BRmetric} see Ref.\citelow{BiKofNL}) of the C-metric leads to the simple classical potential of uniformly accelerated point particle with mass $m$ (for similar results for other BR symmetric spacetimes see Ref.\citelow{BiKofNL}):
$$\Phi = -\frac{Gm}{\sqrt{\rho^2+\left( \zeta-\frac{1}{2}gt^2 \right)^2}}\,.$$

For charged C-metric \re{rCMr} global coordinates appropriate for a Newtonian limit are not known. Therefore we have to introduce suitable class of observers (``static observers at infinity'') -- using purely geometric formulation of the Ehlers frame theory --  and calculate the limit with respect to these observers.

The Newtonian limit then leads to charged massive point particle with standard Newtonian potential and the electromagnetic field according to the Newtonian limit of Maxwell equations as introduced by Kunzle\cite{Kunzle}.

\section{Flat-space limit of the charged rotating C-metric}\label{chap:acc}

In Ref.\citelow{BiKofAcc} we give a detailed account of the special relativistic limit of charged rotating C-metric.
Taking the limit $G\rightarrow 0$ in the metric \re{rCMr} while keeping $a$, $A$, $q$ constant and choosing constant $K=(1+a^2 A^2)^{-1}$ we arrive at flat spacetime in non-trivial coordinates (in Ref.\citelow{BiKofAcc} we give the explicit transformation formulae from these ``uniformly accelerated'' coordinates to Rindler coordinates.). The form of the electromagnetic potential, the field and its invariants remain the same as in Eq.\re{rA}.
The sources of this field are two rotating uniformly accelerating charged discs which become bent in the global inertial frame. This is a generalization of the Born solution for a point particle and of the ``electromagnetic magic field'' studied by Lynden-Bell\cite{DLB-Elm-PRD}.

%\bibliographystyle{ws-procs975x65}
%\bibliography{kofron}

\begin{thebibliography}{10}

\bibitem{BiSchm}
J.~Bi\v{c}{\'{a}}k and B.~Schmidt, {\em Phys. Rev.~D} {\bf 40}, 1827(1989).

\bibitem{BiE}
J.~Bi\v{c}{\'{a}}k, {Selected Solutions of Einstein's Field Equations: Their
  Role in General Relativity and Astrophysics}, in {\em {Einstein's Field
  Equations and Their Physical Implications, Selected Essays in Honour of
  J{\"{u}}rgen Ehlers, Lect. Notes Phys.}\/},  ed. B.~G. Schmidt (Springer,
  Berlin, 2000).

\bibitem{BiPr}
J.~Bi\v{c}\'ak and V.~Pravda, {\em Phys. Rev.~D} {\bf 60}, 044004(1999).

\bibitem{HT2}
K.~{Hong} and E.~{Teo}, {\em Class. and Quantum Grav.} {\bf 22}, 109(2005).

\bibitem{BiKofAcc}
J.~Bi\v{c}{\'{a}}k and D.~Kofro\v{n}, {\em Gen. Relativ. Gravit.} {\bf 41},
  1981(2009).

\bibitem{HT1}
K.~{Hong} and E.~{Teo}, {\em Class. and Quantum Grav.} {\bf 20}, 3269(2003).

\bibitem{BiKofNL}
J.~Bi\v{c}{\'{a}}k and D.~Kofro\v{n}, {\em Gen. Relativ. Gravit.} {\bf 41},
  153(2009).

\bibitem{Ehl-Els}
J.~Ehlers, Newtonian {L}imit of {G}eneral {R}elativity, in {\em Encyclopedia of
  mathematical physics\/},  eds. J.-P. Francoise, G.~L. Naber and S.~T. Tsou
  (Elsevier, 2006) pp. 503--509.

\bibitem{Kunzle}
H.~K\"{u}nzle, {\em Gen. Relativ. Gravit.} {\bf 7}, 445 (1976).

\bibitem{DLB-Elm-PRD}
D.~{Lynden-Bell}, {\em Phys. Rev.~D} {\bf 70}, 105017(2004).

\end{thebibliography}

\end{document}